\documentstyle[12pt]{article}
\catcode`\@=11

   \renewcommand{\section}%
   {\setcounter{equation}{0}\@startsection {section}{1}{\z@}{-3.5ex plus -1ex
   minus -.2ex}{2.3ex plus .2ex}{\Large\bf}}

\catcode`\@=12

\newfam\eusfam
\def\hexnumber@#1{\ifnum#1<10 \number#1\else
 \ifnum#1=10 A\else\ifnum#1=11 B\else\ifnum#1=12 C\else
 \ifnum#1=13 D\else\ifnum#1=14 E\else\ifnum#1=15 F\fi\fi\fi\fi\fi\fi\fi}

\def\Cl{\ifmmode\let\next\Cl@\else
 \def\next{\errmessage{Use \string\Cl\space only in math mode}}\fi\next}
\def\Cl@#1{{\Cl@@{#1}}}
\def\Cl@@#1{\fam\eusfam#1}

\catcode`\@=\active

\def\C{{\rm\kern.24em \vrule width.02em height1.4ex depth-.05ex \kern-.26emC}}
\def\R{{\rm I\kern-.20em R}}
\def\N{{\rm I\kern-.20em N}}
\def\Z{{\rm Z}\!\!\!{\rm Z}}

\newcommand{\be}{\begin{equation}}
\newcommand{\ee}{\end{equation}}
\newcommand{\bz}{{\bar{z}}}

\newcommand{\slc}{\rm{sl}(2,\C)}
\newcommand{\slr}{\rm{sl}(2,\R)}

\newcommand{\ie}{{\rm i.e.}}
\newcommand{\qslc}{U_q\slc}
\newcommand{\half}{\frac{1}{2}}
\newcommand{\cf}{\varphi_{cl}(z,\bz)}
\newcommand{\bp}{\overline{\partial}}

\newcommand{\qvalue}{\frac{p'}{p}}
\newcommand{\vp}{\phi}
\newcommand{\cp}{\varphi_{cl}}
\newcommand{\ncom}{{\cal R}^{root}_{Inf}}
\newcommand{\gncom}{{\cal R}^{gene}_{Inf}}
\newcommand{\com}{{\cal R}_{Fin}}

\makeatletter
\begin{document}

\begin{titlepage}
% \rightline{YITP/U-95-07}
\rightline{September  1996}
\vspace{1.0cm}
\begin{center}
\LARGE{A Note on Quantum  Liouville Theory via 
Quantum Group} \\[.3em]
\Large ------- An Approach to Strong Coupling Liouville Theory ------- \\[1cm]
\large{  Takashi Suzuki$^\ast$ } \\
\normalsize    Department of Electronics \\             
\normalsize    Hiroshima Institute of Technology            \\
\normalsize       Saeki, Hiroshima 731-51 \\
                                 JAPAN 
\end{center}           

\pagestyle{plain}

\vspace{.3cm}

\begin{abstract}
Quantum Liouville theory is annualized in terms of the infinite 
dimensional representations of $\qslc$ with $q$ a root of unity. 
Making full use of characteristic features of the representations, 
we show that vertex operators in this Liouville theory are 
factorized into {\em classical} vertex operators and 
those which are constructed from the  
finite dimensional representations of $\qslc$. 
We further show explicitly that fusion rules in this model also 
enjoys such a factorization. 
Upon the conjecture that the Liouville action effectively 
decouples into the classical Liouville action and that of a quantum 
theory, correlation functions and transition amplitudes are discussed, 
especially an intimate relation between our model 
and geometric quantization 
of the moduli space of Riemann surfaces is suggested. 
The most important result is that our Liouville theory is 
in the strong coupling region, \ie, the central charge $c_L$ 
satisfies $1<c_L<25$. 
An interpretation of quantum space-time is also given within this 
formulation.  
\end{abstract}

\vspace{3.5cm}
-----------------------

{}$^\ast$ e-mail address: stakashi@cc.it-hiroshima.ac.jp

\end{titlepage}
\makeatother

\newpage

\section{Introduction and Setup}

Liouville theory, classically, has a deep connection to the 
geometry of Riemann surfaces. 
Indeed, the solution to the Liouville equation yields Poincar\'e 
metric on the upper-half-plane or the Poincar\'e 
unit disk on which Riemann surfaces are uniformized \cite{Li}. 
In the light of this fact it is natural to expect that quantum Liouville 
theory gives some insights into quantum geometry of surfaces. 
Physically speaking, this problem is nothing but the quantum gravity 
of $2D$ space-time. 
The appearance of the Liouville theory as a theory of quantum 
gravity, so-called Liouville gravity, 
was first recognized by Polyakov in the study of 
non-critical string theory \cite{Po1}. 
In the string theory embedded in the $D$ dimensional target space, 
the partition function is a function $Z[g]$ of the surface metric $g$,  
and is invariant under the group of diffeomorphism acting on the
metric $g$, while it transforms covariantly under local rescalings  
of the surface metric as, 
\be 
Z[e^\Phi \hat g] = e^{\frac{C}{48\pi}S_L(\Phi:\hat g)}\, Z[\hat g]. 
\label{eq:start}
\ee
Here $S_L(\Phi:\hat g)$ is the Liouville action with the background metric 
$\hat g$, and is written as  

\be
S_L(\Phi:\hat g)=\int_\Sigma d^2z 
\sqrt{\hat g}\left\{ \half 
 \hat{g}^{ab}\partial_a \Phi \partial_b \Phi + \Lambda
e^{\Phi(z,\bz)} + R_{\hat g}\Phi(z,\bz) \right\},  
\label{eq:action}
\ee
where $C$ is the central charge of the total system, 
the string coordinates and the reparametrization ghosts. 
The metric on the Riemann surface $\Sigma$ parameterized by  
the complex coordinate $(z,\bz)$ is given by 
$ds^2=  e^{\Phi(z,\bar{z})} {\hat g}_{z\bz} dz d\bar{z}$.  
We denote by $\Lambda$ the cosmological constant. 
$R_{\hat g}$ is the Gaussian curvature measured with the background 
metric $\hat g$ and is given by 
$R_{\hat g}=-2 {\hat g}^{z\bz}\partial_z \partial_{\bz}\log\,
\sqrt{\hat{g}}$.  
Up to now, a number of works (there are too many works to cite here, 
see, e.g., Refs.\cite{Po2}--\cite{DK} and Ref.\cite{Se} for a review) 
have looked at the quantum Liouville gravity and revealed 
many remarkable results. 

One of the important features of the quantum Liouville theory 
is that it possesses quantum group structure of $U_q\slc$ implicitly. 
Precisely, vertex operators in the theory can be 
expressed in terms of the highest weight representations 
of $U_q\slc$. 
We should now remember that, as in the classical algebra $\slc$, 
there are two kinds of highest weight representations 
of $U_q\slc$, \ie, of finite dimension and of infinite dimension, 
and they are completely different from each other.  
Owing to this fact, one can expect that there are two 
entirely different versions of the quantum Liouville theory.  
In one version either finite or infinite dimensional
representations are well-defined and 
we will show that which type of representations 
appears depends on the charges of the vertex operators. 
The quantum Liouville gravity investigated so far are mainly 
associated with the finite dimensional representations of 
$U_q\slc$\cite{Gerv}. 
In this case, however, we have a strong restriction on the central
charge of the Liouville gravity, $c_L\geq 25$, so-called $D=1$
barrier. 
Such gravity is often called the Liouville gravity 
in the weak coupling regime, or for short, 
weak coupling Liouville gravity.  

On the other hand, one expects that the quantum Liouville gravity 
associated with the infinite dimensional representations  
is completely different from the previous one 
and that it can get rid of the barrier, that is, 
the central charge may be in the region $1 < c_L < 25$.  
The Liouville gravity whose central charge is in this region is often 
called the strong coupling Liouville gravity. 
One of the progresses for the strong coupling Liouville gravity 
has been made in Refs.\cite{ST,Ge}.    
They have shown that, upon using infinite dimensional representations of 
$\qslc$, consistent gravity theories can be constructed  
if and only if the central charge takes the special values 
$7,\, 13$ and $19$. 
This result makes us confident that  the Liouville gravity associated with 
the infinite dimensional  representation is in the strong coupling regime. 
However, as will be explained in Section 3,  
the deformation parameter $q$ of the algebra $\qslc$ 
dealt with in Ref.\cite{Ge} is not a root of unity.  
It should be emphasized here that  the infinite dimensional 
representations of $\qslc$ when $q$ is a root of unity are 
drastically different from those with generic $q$ \cite{MS}. 
In particular, a new and remarkable feature is that 
every irreducible infinite dimensional representations 
necessarily factorizes into a representation of the {\it classical} 
algebra $\slr$ and a {\it finite} dimensional one of $\qslc$. 
Along this line, it is worthwhile to investigate the Liouville gravity
associated with the infinite dimensional representations with $q$ at a
root of unity and observe how the characteristic feature of the 
representations works in the theory of Liouville gravity. 
Motivated by this, the aim of this paper is to investigate 
the quantum Liouville gravity via such representations 
with hoping that such a theory would lead us to a different 
strong coupling Liouville gravity from \cite{Ge}. 

Hereafter, let us use the notations, $\ncom$ and $\gncom$ 
the infinite dimensional representations of $\qslc$ 
{\it with $q$ at a root of unity} and {\it generic} $q$, respectively, 
and $\com$ the finite dimensional ones. 

We will see that the fields $\Phi(z,\bz)$ of the Liouville theory 
based on $\ncom$ should be expanded around the classical 
Liouville field $\cf$ as 
$\Phi = \cp \oplus \phi$, $\phi$ representing quantum fluctuations 
around the classical field (see also Ref.\cite{TS}). 
In other words, the metric with which our quantum Liouville theory is 
constructed should be chosen as $ds^2=e^{\kappa\phi(z,\bz)} \hat g_P$ 
where $\kappa$ is some constant and $\hat g_P$ is the Poincar\'e metric 
$\hat g_P =e^{\cf}dz d\bz$ playing the role of the background metric. 
In view of this, it will turn out that our Liouville theory is effectively 
composed of two Liouville theories, one is the classical Liouville 
theory $S^{cl}(\cp)$ and the other is the quantum Liouville theory 
$S^q(\phi:\hat g_P)$ with respect to the field $\phi(z,\bz)$. 
The sector $S^q(\phi:\hat g_P)$ represents quantum fluctuations 
around $S^{cl}(\cp)$ and is the Liouville theory associated with $\com$. 
We will explicitly show that our Liouville gravity  
is actually in the  strongly coupled regime, \ie, $1<c_L<25$.  

% Although the explicit relation between the quantum Liouville gravity 
% of this paper and that of Takhtajan is yet to be discussed, 
% one can expect an intimate relation between the two. 

The organization of this paper is as follows: 
Section 2 looks at the classical Liouville theory in some
details, especially from the viewpoints of geometric and algebraic 
structures of the theory. 
The discussions of the quantum Liouville gravity 
are given in Section 3.  
We will review in \S 3-1 the weak coupling Liouville theory briefly, 
and other sections, \S 3.2--\S 3.5 are devoted to our main concern,  
\ie, the Liouville theory based on $\ncom$.   
We will show the decomposition of our theory into the 
classical theory and the quantum one,  
and give a concept of {\em quantum space-time}. 
We further discuss about fusion rules and transition amplitude. 
 It will turn out that our formulation of the quantum Liouville 
theory admits a nice interpretation in the context of the 
geometric quantization of K\"ahler geometry of moduli space. 
Here the classical sector, whose appearance is the peculiarity of 
our formulation, of correlation function plays an important role, 
\ie, the Hermitian metric of a line bundle over the moduli space. 
In section 4, some discussions are given. 
Appendix reviews  the infinite dimensional representations of 
$\qslc$ with $q$ at a root of unity.

%%%%%%%%%%%%%%%%%%%%%%%%%%%%%%%%%%%%%%%%%%%%%%%%%%%%%%%%%%%%%%%%%%%%%%%%%%%%%%

\section{Classical Liouville Theory} 

In this section, we will introduce some backgrounds of the classical 
Liouville theory and Riemann surfaces for our later use, 
especially algebraic and geometric aspects of the theory. 

It is known that the classical Liouville theory describes 
hyperbolic geometry of Riemann surfaces. 
Let $\Sigma_{g,N}$ be a Riemann surface with genus $g$ and 
$N$ branch points $\{z_i\}$ of orders $k_i\in \rm{I}\!\rm{N}_{\geq 2}$. 
The equation $\delta S_L=0$ yields the Liouville equation 
of motion 
\be 
\partial \bp \Phi(z,\bz) = \frac{\Lambda}{2}
 e^{\Phi(z,\bz)}. 
\label{eq:equation}
\ee
Here we have made a specific choice of the background metric as 
${\hat g}_{ab}=\delta_{z\bz}$. 
In that case, the Gaussian curvature is given by 
$R_g= -\Lambda$, \ie, 
constant curvature. 
According to the Gauss-Bonnet theorem, 
$\Sigma_{g,N}$ admits metrics $\,g\,$ with constant negative curvature, called 
the Poincar\'e metric, 
for the negative Euler characteristic $\chi(\Sigma_{g,N})<0$.  
Noticing that, for any metric $\tilde g$ on $\Sigma_{g,N}$, there 
exists a scaling factor $e^\lambda$ such that 
$g= e^{-\lambda}\tilde g$ has constant negative curvature 
$R_g=-1$, we will set $\Lambda=1$ hereafter. 
Note however that such a setting cannot be allowed for the 
quantum case. 

Let us explain the connection between the Liouville theory and 
the Fuchsian uniformization of Riemann surfaces. 
The uniformization theorem states that every Riemann surface 
with negative constant curvature is conformally equivalent to 
the quotient of the unit disk $D=\{w \in \C \vert 
\, \vert w\vert <1\}$ by the action of a finitely generated 
Fuchsian group $\Gamma$, \ie, $\Sigma_{g,N} \cong D/\Gamma$. 
In terms of the uniformization map, 
$J_\Sigma\,:\, D \rightarrow \Sigma_{g,N}$ 
a solution to the Liouville equation (\ref{eq:equation}) 
is written as  
\be 
e^{\cf} = 
\frac{\vert \partial_z J_\Sigma^{-1}(z)\vert^2}
{(1- \vert J_{\Sigma}^{-1}(z){\vert^2}){}^2}. 
\label{eq:csolution}
\ee    
Upon rewriting $J_\Sigma^{-1}(z) = w$, the coordinate on $D$, 
the solution (\ref{eq:csolution}) gives the Poincar\'e metric on $D$, 
\be 
ds^2 = \frac{dw \wedge d\bar{w}}{(1-w\bar{w})^2}. 
\ee
Note that the $PSL(2,\rm{I}\!\rm{R})$ fractional transformations 
for $J_\Sigma^{-1}(z)$ leave the metric invariant. 
% This is the evidence of the fact that the classical Liouville theory 
% possesses the $\SLC$ symmetry. 

The classical Liouville theory is a conformally invariant theory. 
This is due to the fact that the energy-momentum (EM) tensor defined 
by $T_{ab}=2\pi/\sqrt{\hat g}(\delta S_L/\delta\hat{g}_{ab})$ 
is traceless, \ie, $T^{cl}_{z\bz}=0$. 
The $(2,0)$-component is given by 
\be
T^{cl}_{zz}=\frac{1}{\gamma^2}\left[ -\half \partial_z\cp\partial_z\cp 
       + \partial_z^2\cp \right], \label{eq:emtensor} 
\ee 
and is conserved, \ie, $\partial_\bz T^{cl}_{zz}=0$. 
Here it should be emphasized that the second term in 
eq.(\ref{eq:emtensor}) coming from the last term of the action 
(\ref{eq:action}) is indispensable in order for the EM-tensor 
to be traceless and satisfy the correct transformation law of a 
projective connection under holomorphic coordinate change 
$z=f(\tilde z)$, 
\begin{eqnarray}
\tilde T^{cl}_{\tilde z \tilde z}(d\tilde z)^2 
&=& \left(\frac{df}{d\tilde z}\right)^2 T^{cl}_{zz} (dz)^2 + 
\frac{1}{\gamma^2}\{f,\tilde z\}_S (d\tilde z)^2, \nonumber \\
\{f,\tilde z\}_S &=& \frac{f'''}{f'}
-\frac{3}{2}\left(\frac{f''}{f'}\right)^2.  
\label{eq:projective}
\end{eqnarray}
Indeed, using the fact that the metric $e^{\Phi(z,\bz)}$ 
is a primary field of dimension $(1,1)$, or equivalently, 
it transforms under the coordinate change as 
\be
e^{\tilde\cp(\tilde z,\bar{\tilde z})}d\tilde z d\bar{\tilde z} 
=\left\vert\, \frac{d\tilde z}{dz}\,\right\vert^2 \,e^{\cp(z,\bz)}dz d\bz,
\ee 
one can check that the EM-tensor satisfies 
the transformation law (\ref{eq:projective}). 
This transformation law of EM-tensor indicates that the 
central charge $c_0$ of the classical theory is 
\be 
c_0=\frac{12}{\gamma^2}. \label{eq:classicalcentral}
\ee

We next discuss the algebraic structure of the classical Liouville 
theory. 
To see this, let us take the $\alpha$-th 
power of the metric (\ref{eq:csolution}) at a special point $P$. 
Geometrically it corresponds to a branch point $P$ on the surface, and 
$\alpha$ is related to the order $k$ of the branch point as 
\be 
\alpha_i=\frac{1-k^{-2}}{2\gamma^2}. 
\label{charge}
\ee
One immediately finds there are two entirely different 
regions of $\alpha$, namely, 
(I) $\alpha=-j\leq0,\, (j\in \Z/2)$ and  (II) $\alpha=h>0$. 
It lies in this fact that there are two kinds of quantum Liouville 
gravity, \ie, of strong coupling and weak coupling regimes. 
In order to see this more explicitly, 
we calculate the $\alpha$-th power of eq.(\ref{eq:csolution}),  
\begin{eqnarray}
({\rm I}) \quad &e^{-j\cp(z,\bz)}& = 
\left(\frac{1-\vert J_\Sigma^{-1}\vert^2}
{\sqrt{\partial_z J_\Sigma^{-1} \partial_{\bz}
\overline{J_\Sigma^{-1}}}}
\right)^{2j}=
\sum_{m=-j}^j N^j_m \psi^j_m(z)\psi^{j,m}(\bz)  \label{eq:I} \\
({\rm II}) \quad &e^{h\cp(z,\bz)}& = 
\left(\frac{\sqrt{\partial_z J_\Sigma^{-1} \partial_\bz J_\Sigma^{-1}}}
{1-\vert J_\Sigma^{-1}\vert^2}
\right)^{2h}= \sum_{r=0}^\infty N^h_r \lambda^h_r(z)\lambda^{h,r}(\bz), 
\label{eq:II}
\end{eqnarray}
where $N^j_m,\,N^h_r$ are binomial coefficients. 
Although we can represent the functions $\psi^j_m(z)$ and 
$\lambda^h_k(z)$ in terms of a free field, we are not interested 
in the explicit expressions within the latter discussions. 
The crucial fact is that $\psi^j_m(z)$ and $\lambda^h_r(z)$ form, 
respectively, the finite and the infinite dimensional representations,  
$V^{cl}_j$ and $V^{cl}_h$, of $\slc$:  
Denoting by  $E_+, E_-, H$ the generators of $\slc$ 
satisfying the relations $[E_+,E_-]=2H, [H, E_\pm]=\pm E_\pm$, 
these representations are, 
\begin{eqnarray}
V^{cl}_j &=& \{ \psi^j_m(z) \, \vert\, E_+\psi^j_j = E_-\psi^j_{-j}=0, 
H\psi^j_m=m\psi^j_m, \, -j\leq m \leq j \}, \nonumber \\ [.1cm] 
V^{cl}_h &=& \{ \lambda^h_r (z)  \,\vert\,  E_-\lambda^h_0=0, 
H\lambda^h_r =(h+r) \lambda^h_r, \, r=0, 1, \cdots \}.  
\end{eqnarray} 
One can further show that the chiral sector $\psi^j_m(z)$ and  
$\lambda^h_r(z)$ satisfy the Poisson-Lie relations of the algebra
$\slc$\cite{GN,FT}. 
\be
\{\psi^{j_1}_{m_1}(z_1)\stackrel{\otimes}{,}\psi^{j_2}_{m_2}(z_2)\}
=-\pi\gamma^2(r^{j_1j_2})^{m_1'm_2'}_{m_1m_2}
\psi^{j_2}_{m'_2}(z_2)\psi^{j_1}_{m'_1}(z_1), \label{eq:poisson}
\ee
where $r^{j_1j_2}$ is the $\slc\otimes\slc$ valued 
classical $r$-matrix, $r=H\otimes H+E_+\otimes E_-$. 
The same relations hold also for $\lambda^h_r(z)$. 
Upon the general philosophy that 
the quantization of the Poisson-Lie algebra ${\bf {\rm g}}$ 
yields the quantum universal enveloping algebra $U_q{\bf {\rm g}}$ 
endowed with Yang-Baxter relation, 
these algebraic structures will be essential later.  

As the last comment, it is necessary for the later discussions to 
summarize the geometrical aspects of the classical Liouville action. 
Since our main concern will be on the $N$ punctured 
sphere, \ie, a sphere with $N$ branch points of orders infinity, 
we will confine ourselves to the surface $\Sigma_{0, N}$ where 
punctures are located at $\{z_1, z_2, \cdots, z_N=\infty\}$. 
Two Riemann surfaces  of this type are isomorphic if and only if 
they are related by an element of the group $PSL(2,\C)$ $-$ 
a group of all automorphisms of ${\bf P}^1$. 
Using this freedom we can normalize such Riemann surfaces by setting 
$z_{N-2}=0, z_{N-1}=1, z_N=\infty$, then 
$\Sigma_{0,N}=\C\backslash \{z_1, \cdots, z_{N-3}, 0, 1\}$. 
Defining the space of punctures as 
${\cal T}_N=\{(z_1, \cdots, z_{N-3})\in \C^{\otimes N-3}
\vert z_i\neq z_j, \,  {\rm for}\, i\neq j\}$, 
one sees that a point in ${\cal T}_N$ represents a Riemann surface 
of the type $\Sigma_{0,N}$. 
% ${\cal T}_N$ is nothing but the Teichm\"uler space of the surfaces 
% $\Sigma_{0,N}$ and ${\rm dim}\,{\cal T}_N= N-3$. 
Moreover, if two $\Sigma_{0,N}$'s are connected by an action of the
symmetric group ${\rm Symm}(N)$, 
they should be regarded as the same with each other. 
Hence, we get the moduli space of Riemann surfaces of the type 
$\Sigma_{0,N}$ as,  
\be
{\cal M}_{0,N} = {\cal T}_N / {\rm Symm}(N). \label{eq:moduli}
\ee

For the surface $\Sigma_{0,N}$, the Liouville field $\cp(z,\bz)$ 
has the following asymptotics near the punctures, 
\be 
\cp(z,\bz) \stackrel{z\rightarrow z_i}{\longrightarrow} \left\{ 
\begin{array}{ll} 
-2\log\,\epsilon_i -2\log \vert \log\,\epsilon_i\vert,  \quad\,\, 
\epsilon_1\equiv\vert z-z_i\vert, \quad & {\rm for}\quad i\neq N, \\[.15cm]
-2\log\,\vert z\vert -2\log\,\vert \log\vert z\vert\vert, 
\quad & {\rm for} \quad i=N.  
\end{array}\right. \label{eq:asymp}
\ee 
Of course, such asymptotics enjoy the Liouville equation 
(\ref{eq:equation}). 
Denote by $\Im$ a class of fields on $\Sigma_{0,N}$ satisfying
asymptotics (\ref{eq:asymp}). 
Due to the asymptotics, the action (\ref{eq:action}) diverges 
for $\cp\in \Im$ and a regularized action has been obtained by 
Takhtajan and Zograf \cite{ZT} in a reparametrization invariant 
manner as 
\be
\overline{S}_L(\cp) =\lim_{\epsilon\rightarrow 0}
\left\{
\int_{\Sigma_\epsilon} d^2z\left( \partial_z\cp\partial_{\bz}\cp
+ e^{\cp(z,\bz)}\right)+ 2\pi N\log\,\epsilon+4\pi(N-2)
\log\vert\log\,\epsilon\vert\right\}, 
\label{eq:Action}
\ee
where $\Sigma_\epsilon=\Sigma\setminus \cup_{i=1}^{N-1}
\{\vert z-z_i\vert <\epsilon\}\cup\{\vert z\vert >1/\epsilon\}$. 
The Euler-Lagrange equation $\delta \overline{S}_L =0$ derives 
again the Liouville equation (\ref{eq:equation}). 
Owing to the regularization, however, the classical action 
$\overline{S}_L(\cp)$ is no longer invariant under the action of 
${\rm Symm}(N)$ on ${\cal T}_N$. 
According to Ref.\cite{Zo}, one can find, for a real constant $k$, 
1-cocycle $f^k_\sigma$ of ${\rm Symm}(N)$ satisfying 
\be
\exp\left(\frac{k}{\pi}\overline{S}_L(\cp)\circ \sigma\right)
\vert f^k_\sigma\vert^2 =
\exp\left(\frac{k}{\pi}\overline{S}_L(\cp)\right). 
\label{eq:moduleranomaly}
\ee
Here ${\rm Symm}(N)$ acts on the trivial bundle 
${\cal T}_N \times {\C} \rightarrow {\cal T}_N$ as 
$(t, z)\mapsto (\sigma t, f^k_\sigma(t) z)$, where 
$ t\in {\cal T}_N,\, z\in {\C}$ and $\sigma\in{\rm Symm}(N)$.  
It follows that the cocycle $f^k_\sigma$ defines a Hermitian 
line bundle ${\cal L}_k={\cal T}_N \times {\C} / {\rm Symm}(N)$ 
over the moduli space ${\cal M}_{0,N}$ and the function 
$\exp(\frac{k}{\pi}\overline{S}_L^{\cp})$ can be interpreted as a 
Hermitian metric in the line bundle ${\cal L}_k$. 
This fact will play an important role in the next section.

%%%%%%%%%%%%%%%%%%%%%%%%%%%%%%%%%%%%%%%%%%%%%%%%%%%%%%%%%%%%%%%%%%%%%%%

\section{Quantum Liouville Gravity}

As we have seen in the previous section, the classical 
Liouville theory describes the hyperbolic geometry of 
Riemann surfaces $\Sigma$. 
Indeed, the classical Liouville field $\cf$  
is a function on $\Sigma$ defining  the Poincar\'e  metric 
in terms of the Fuchsian uniformization map. 
On the contrary, in quantum theory, the Liouville field 
$\Phi(z,\bz)$ expresses quantum fluctuations of the metric 
and so does the uniformization map $J_\Sigma^{-1}(z)$.  
In other words, the coordinates $(w, \bar{w})$ on the unit disk 
are not ordinary complex numbers but in some sense quantum objects. 
% In this fact, we can glance at quantum fluctuations of 
% metrics of the surface. 
This is the reason why we regard the quantum Liouville theory as 
a theory of  quantum 2$D$ gravity or quantum geometry of surfaces. 

\subsection{Introduction of Quantum Liouville Action}

We start by summarizing how the quantum Liouville theory appears 
in the string theory. 
The original definition of the string partition function is 
\be
Z=\int d\tau\, [d\Phi]_g[dX]_g [d(gh)]_g
\large{e^{-S_X(X;g)-S_{gh}(b,c;g)-\Lambda_0\int\sqrt{g}}},
\label{string}
\ee
with $\tau$ the Teichm\"uler parameter, $X=\{X^\mu\}$ the string 
coordinates embedded in the $D$ dimensional target space and 
$(gh)$ stands for the ghost coordinates $\{b,c\}$ associated with 
the diffeomorphism invariance. 
Choosing a metric slice $g=e^{\Phi}\hat{g}$ gives the following relation 
for the path integral measures, 
\be
[d\Phi]_{e^{\Phi}\hat{g}}[dX]_{e^{\Phi}\hat{g}} [d(gh)]_{e^{\Phi}\hat{g}}
=J(\Phi;{\hat{g}})\, 
[d\Phi]_{\hat{g}}[dX]_{\hat{g}}[d(gh)]_{\hat{g}}
\ee
where $J(\Phi;{\hat{g}})$ is the Jacobian. 
The contributions to the Jacobian from $[dX]$ and $[d(gh)]$ was 
obtained by Polyakov \cite{Po1}, and that from 
$[d\Phi]$ was postulated \cite{DK} so that the partition function 
(\ref{string}) finally had the following form,  
\begin{eqnarray}
Z &=& \int d\tau\, [d\Phi]_{\hat{g}}[dX]_{\hat{g}} [d(gh)]_{\hat{g}}\,
\large{e^{-S_X(X;\hat{g})-S_{gh}(b,c;\hat{g})} }
\label{string2} \\[.15cm] 
 &&\qquad\qquad\qquad\quad\times\, 
\large{e^{-\int d^2\xi \sqrt{\hat{g}}
\left( \frac{A}{2}\hat{g}^{ab}
\partial_a\Phi \partial_b\Phi + BR_{\hat{g}}\Phi +\Lambda 
e^{C\Phi}\right)}}. \nonumber
\end{eqnarray}
The constants $A, B$ are determined by calculating the responses 
of the Weyl rescaling $\hat{g}\,\rightarrow\,e^{\sigma}\hat{g}$ and 
demanding the invariance of $Z$, one finds $A=B=\frac{25-D}{48\pi}$. 
Upon replacing $\Phi\,\rightarrow\sqrt{\frac{12}{25-D}}\Phi$, 
we obtain the Liouville action 
\be
\frac{1}{4\pi}S_L(\Phi:\hat g)=\frac{1}{4\pi}
\int_\Sigma d^2\xi \sqrt{\hat g}\left\{ \half
 \hat{g}^{ab}\partial_a \Phi \partial_b \Phi
+ Q_0R_{\hat g}\Phi+ \Lambda e^{\gamma\Phi}
 \right\},  \label{waction}
\ee
where 
\be
Q_0=\sqrt{\frac{25-D}{12}}. 
\ee
On the other hand, the constant $C$ plays the role of 
renormalized coupling constant. 
In eq.(\ref{waction}), we have replaced $OCC$ by $\gamma$, 
since the field $\Phi$ has been rescaled.  and $\gamma$ 
will be related to the deformation parameter $q$ of quantum group 
$\qslc$. 

% This Liouville action is invariant under the Weyl transformation 
% together with the shift of the Liouville field; 
% \be
% \hat{g}\,\rightarrow\,e^\sigma \hat{g}, \quad 
% \gamma \Phi\,\rightarrow\,\gamma \Phi-Q_0\sigma. \label{Weyl}
% \ee
% Due to the invariance, we can freely choose the background metric. 

The energy-momentum  tensor of the Liouville sector is obtained  as  
\be
T_{zz} = 
-\half (\partial_z \Phi)^2 +  Q_0 \partial^2_z\Phi,
\quad\, T_{z\bz}=0. 
\label{eq:qEM}
\ee
The second equation is derived upon the equation of motion. 
The central charge of the quantum Liouville theory is calculated as 
\be
c_L=1+12Q_0^2.   \label{eq:central}
\ee
Notice that the total conformal anomaly vanishes, 
\be
c_X + c_L +c_{gh} =D+(26-D)-26=0,
\ee
which is consistent with the requirement of the conformal invariance. 

For the time being, we will concentrate only on the Liouville sector. 
The correlation function of $N$ vertices are given by 
\be
\left\langle V_{\alpha_1}(z_1,\bz_1) V_{\alpha_2}(z_2,\bz_2) \cdots 
V_{\alpha_N}(z_N,\bz_N) 
\right\rangle  = \int_{\Im} [d\Phi] \, 
e^{-\frac{1}{4\pi}S_L(\Phi:\hat{g})}\,
\prod_{i=1}^N V_{\alpha_i}(z_i,\bz_i), 
\label{eq:correlation1}
\ee
where the Liouville vertex operator with charge $\alpha_i$ is given by 
\be
V_{\alpha_i}(z_i,\bz_i)=e^{\alpha_i\gamma\Phi(z_1,\bz_i)}. 
\ee
Notice that, in this definition of the correlation function, 
the base manifold on which the vertices live is not 
a manifold with  branch points but just the sphere ${\bf P}^1$. 
Then the functional integral is performed over the space 
$\Im$ of all smooth metrics $\Phi(z,\bz)$ on ${\bf P}^1$. 
Since puncture corresponds to the branch point of order infinity, 
all the punctures on $\Sigma_{0,N}$ correspond to the vertex operators
with charges $\alpha_i=1/2\gamma^2$ (see eq.(\ref{charge})). 
As in the classical theory,  (\ref{eq:I}) and (\ref{eq:II}), 
there are two distinct regions according to the value of the charge 
$\alpha$, \ie, (I) $\alpha\leq 0$ and (II) $\alpha>0$. 
We will refer these two regions as (I) weak coupling region and 
(II) strong coupling region.
\footnote{According to the standard convention, 
the Liouville theories which are defined in the spacetime of dimension 
$D\leq 1$ and $1<D<25$ are called, respectively, weak coupling 
theory and string coupling theory. 
The reason why we use here the terms weak and strong is that 
it will be turned out later that vertices defined 
in the weak (strong) coupling theory carry negative(positive) charges.} 
In quantum theory, however, the difference between them becomes 
more sever than the classical theory. 
There is a big gap known as the $D=1$ barrier between two regions.  
The weak coupling region, studied first by KPZ and DDK,  
has been investigated by many authors.  
On the other hand, the quantum Liouville theory in the strong 
coupling region is a long-standing problem. 
Our concern in this article is on this region and 
we will investigate it by using quantum group methods. 
Before doing this, it is instructive to review the Liouville 
theory in the weak coupling region.

\subsection{Brief Review of the Weak Coupling Region}

The standard approach to the weak coupling Liouville theory 
begins  with the action (\ref{waction}). 
Till now remarkable progresses have been done by many authors.   
Let us, in this section, observe  
the quantum group aspects of the weak coupling Liouville theory. 
The quantum group structure appears through the vertex operators, 
and has been studied extensively in \cite{Gerv,ST}. 
A crucial fact is that, in this region,  
the finite dimensional representations of $\qslc$ appear.  
To be precise, writing the vertex operator with charge $-j$ as 
\be 
e^{-j\gamma\Phi(z,\bz)}=\sum_{m=-j}^j {\cal N}^j_m 
\Psi^j_m(z)\overline{\Psi}^{j,m}(\bz), 
\ee
they have shown that $\Psi^j_m(z),\, m=-j, \cdots, j$ 
form the $2j+1$ dimensional representation of $\qslc$ with 
\be
q=e^{\pi i \frac{\gamma^2}{2}},  \label{defq}
\ee 
and satisfy the braiding-commutation relations, 
\be
\Psi^{j_1}_{m_1}(z_1)\otimes \Psi^{j_2}_{m_2}(z_2)
= \left(R^{j_1 j_2}\right)^{m_1' m_2'}_{m_1 m_2}
\Psi^{j_2}_{m'_2}(z_2)\otimes \Psi^{j_1}_{m'_1}(z_1), 
\label{eq:braid}
\ee
where $R^{j_1j_2}$ is the universal $R$-matrix of $\qslc$. 
The factorization property of the quantum vertices is a natural guess 
from the classical result (\ref{eq:I}). 
The braiding-commutation relation (\ref{eq:braid}) 
reduces to the Poisson-Lie relation (\ref{eq:poisson}) 
in the classical limit, $\gamma\rightarrow0$. 
This fact is in agreement with the general concept of quantum groups
as the quantization of Poisson-Lie algebras of classical Lie groups. 

An important fact of the region (I) arises in 
the relation between $Q_0$ and the renormalized coupling constant $\gamma$,  
that is, 
\be
Q_0=\frac{1}{\gamma} +\frac{\gamma}{2}.  \label{eq:QW}  
\ee
% This relation is derived by calculating anormalous dimension of a 
% vertex operator. 
% It is well-known that anormalous dimension can be obtained by 
% analizing braiding relation, which is deeply related to 
% the $q$-$6j$ symmbol. 
% In other words, the relation (\ref{eq:QW}) can be obtained 
% only from quantum group structure. 
Substituting (\ref{eq:QW}) into the general expression of the 
central charge (\ref{eq:central}), one obtains 
\be
c_L=13+6\left(\frac{\gamma^2}{2} + \frac{2}{\gamma^2}\right)  
\label{eq:weakcentral}
\ee
and finds that the central charge always satisfies $c_L\geq 25\,(\leq 1)$ 
for a real (imaginary) $\gamma$. 
Notice that $\gamma\Phi$ should be real since $e^{\gamma\Phi}$ is 
the metric in this model. 
The very origin that restricts the theory to the weak coupling 
region is 
the relation (\ref{eq:QW}). 
This was first obtained 
in Ref.\cite{CT} by requiring that the EM tensor (\ref{eq:qEM}) 
satisfies Virasoro algebra on the cylindrical basis $S^1\times \R$.  
Once Virasoro structure is found, one can apply the Coulomb gas 
formulation of the minimal CFT to the Liouville theory, 
although the Liouville field is no longer a free field. 
Let us explain briefly. 
Notice  that the parameter $Q_0$ corresponds to $i\alpha_0$, 
$\alpha_0$ being the background charge in the Coulomb gas formulation 
and the highest weight vector $\Psi^j_j(z)$ can be related to  the
primary field of the type $V_{2j+1, 1}(z)$ in the Kac's table, 
where $V_{n,m}(z)= :e^{i\alpha_{n,m}\hat\phi(z)}:$ with the charge 
$\alpha_{n,m}=\{(1-n)\alpha_++(1-m)\alpha_-\}/2$ and a free field 
$\hat\phi(z)$.  
Upon comparing eq.(\ref{eq:QW}) and the relations 
$\alpha_0=(\alpha_+ +\alpha_-)/2,\, \alpha_+\alpha_-=-2$, 
one immediately sees that the renormalized coupling constant 
$\gamma$ is related to one of the screening charges, 
say, $\alpha_+$ as $\gamma=i\alpha_+$.  
With the above connections at hand, 
the correspondence between the metric $e^{\gamma\Phi(z,\bz)}$ and 
the screening operator $e^{i\alpha_+\hat\phi(z)}$ which has 
conformal dimension $1$ is now clear. 
Thus we obtain the correct dimension $(1,1)$ for  the metric 
$e^{\gamma\Phi(z,\bz)}$ \cite{DK}. 
Furthermore, the relation (\ref{eq:QW}) is  consistent with 
the quantum group structure of $\qslc$ when the representation is 
of finite dimension. 
Indeed, the braiding commutation relation which is deeply related 
to the $q$-$6j$ symbols indicates that the operator $e^{-j\gamma\Phi}$ 
has dimension $-j-\frac{\gamma^2}{2}j(j+1)$. 
By comparing with the result of CFT that the dimension of the operator 
is given as $-\frac{1}{2}j\gamma(j\gamma+2Q_0)$, 
we then obtain (\ref{eq:QW}). 
Thus in the weak coupling Liouville theory, the relation (\ref{eq:QW}) 
works well and plays an important role, although 
it is the origin of the severe problem, the $D=1$ barrier. 
%%%%%%%%%%%%%%%%%%%%%%%%%%%%%%%%%%%%%%%%%%%%%%%%%%%%%%%%%%%%%%%%
%
%                         3.1
%
%%%%%%%%%%%%%%%%%%%%%%%%%%%%%%%%%%%%%%%%%%%%%%%%%%%%%%%%%%%%%%%%

\subsection{The Strong Coupling Region}
Now we are at the stage to enter into  the mysterious world 
of the strong coupling Liouville theory. 
At the beginning, we propose a Liouville action for the 
strong coupling region, 
\be
\widehat{S}_L(\Phi:\hat g) = S_L +S', \quad\, \mbox{with} \quad\, 
S'=\lambda \int d^2z \sqrt{\hat  g}e^{\cp(z,\bz)}, 
\label{eq:qaction}
\ee
where $S_L$ is the Liouville action (\ref{waction}).  
We have introduced the additional cosmological term $S'$ whose necessity 
will be  understood later. 
Note here that, owing to this term, 
the potential reduces to the classical metric $e^{\cp(z,\bz)}$ 
in the limit $\Phi\rightarrow -\infty$, while, without this term, 
the potential vanishes in the limit.  
% The holomorphic sector of the energy-momentum tensor is exactly same as 
% eq.(\ref{eq:qEM}), and so  
% the central charge is also given by eq.(\ref{eq:central}). 
% on the other hand, the trace part does not disappear even uopn the 
% equation of motion for $\Phi$. 
% Owing to the additional cosmological term $S_1$, we have 
% \be 
% T_{z\bz} = \Lambda_1 e^{\cp(z,\bz)}.
% \ee
% The central charge is also given by eq.(\ref{eq:central}). 
% Moreover the action is not invariant under the Weyl rescaling 
% with $\Phi$ translation (\ref{Weyl}). 
% It recovers if the classical field $\cp$ is shifted 
% $\cp\rightarrow\cp-\sigma$ simltaneously.  
We will work, however, not  with this action 
but with the representations of the quantum group $\qslc$ 
in the following discussions. 

Let us turn to the quantum group structure. 
To do this, we introduce vertex operator $e^{h\gamma\Phi(z,\bz)}$   
denoting the charge $h$ instead of $\alpha$ in this region. 
By definition the charge $h$ is always positive. 
As in the weak coupling case, the vertex operator allows 
holomorphic decomposition as, 
\be
e^{h\gamma\Phi(z,\bz)}=\sum_{r=0}^\infty {\cal N}_q^{h,r} \Lambda^h_r(z) 
\overline{\Lambda}^{h,r}(\bz). \label{eq:strongvertex}
\ee  
It can be shown that the vectors $\Lambda^h_r(z), r=0, 1, \cdots$ 
form an infinite dimensional highest weight representation 
$V_h$ of $\qslc$ with (\ref{defq}). 
Thus in this region, infinite dimensional  representations 
appear instead of finite ones in the weak-coupling region. 
The question here is the relation between $Q_0$ and the 
renormalized coupling constant $\gamma$. 
If we require $c_L>1$ with maintaining the relation 
(\ref{eq:QW}), the constant $\gamma$ becomes complex. 
This fact means that, as suggested in the previous section 3.1, 
the screening charges become complex and, therefore, 
the dimensions of primaries are not real in general. 
Gervais and his collaborators \cite{Ge} have 
discussed the strong coupling Liouville theory 
under the situation. 
Their solution to the problem is summarized in \lq\lq 
truncation theorem'', which guarantees that 
only in the special central charges $c_L=7, 13$ and $19$, 
the chiral components of the Liouville vertices form 
Virasoro Verma modules with real highest weights.  
Upon the relation (\ref{eq:QW}) together with the central charge  
(\ref{eq:central}), these central charges 
$c_L=7, 13$ and $19$ give the values of the constants $\gamma$, 
respectively, $\gamma^2=-1+i\sqrt{3},\, 2i$ and $\gamma^2=1+i\sqrt3$.  
Notice here that, with the definition eq.(\ref{defq}), 
the parameter $q$ of the quantum group $\qslc$ is generic. 
In such a situation the relation (\ref{eq:QW}) is actually consistent. 
As in the weak coupling region, the braiding commutation relation 
in the case of infinite dimensional representations 
shows that the dimension of the operator $e^{h\gamma\Phi}$ 
is calculated as $h-\frac{\gamma^2}{2}h(h-1)$. 
By comparing this dimension with that obtained from the CFT method, 
that is, $-\frac{1}{2}h{\gamma}(h\gamma-2Q_0)$, we again obtain 
(\ref{eq:QW}). 

Now we would like to ask a question, what will happen if 
we take the parameter $q$ being a root of unity. 
With the idea that the infinite dimensional representations 
$\ncom$ and $gncom$ are drastically different from each other, 
it is quite natural to expect that the quantum Liouville theory 
associated with $\ncom$ has big difference from that based on $\gncom$. 
Furthermore, since the $q$-$6j$ symbols for the representations $\ncom$ 
is completely different from those for $\gncom$, 
we could get rid of the relation (\ref{eq:QW}) and, therefore,  
have any other values of the central charge. 
In the following discussions, 
the only essential assumption  is that our Liouville theory 
possesses the structure of $\ncom$. 

Let $q$ be a $p$-th root of unity, 
\footnote{ 
In this paper, as a matter of convention, 
$q$ such that $q^p=-1$ for ${}^\exists p\in \N$ 
is also called a root of unity. }, 
\ie,   
\be
 q=e^{\pi i \qvalue} \quad \quad {\rm \ie}, \quad 
\frac{\gamma^2}{2}=\qvalue, \label{eq:gvalue}
\ee
with $p, p'\in \N$, coprime with each other. 
Such representations are reported in the Appendix. 
Once the parameter $q$ is set to the value (\ref{eq:gvalue}), 
the infinite dimensional 
highest weight representations are parameterized by two integers 
$\mu$ and $\nu$ such that $h\rightarrow h_{\mu\nu}=\zeta p-j$ where 
$\,\zeta=\nu/2p', \,j=(\mu-1)/2$. 
We denote by $V_{\mu,\nu}$ the module on the highest weight
state of weight $h_{\mu\nu}$ and by $\Lambda^{\mu\nu}_r(z)$ 
the $r$-th weight vector in the module $V_{\mu,\nu}$, namely,  
it corresponds to the $r$-th holomorphic sector  
in the decomposition of the vertex operator $e^{h_{\mu\nu}\Phi(z,\bz)}$. 
One immediately sees that the terms $\Lambda^{\mu\nu}_{kp+\tilde s}(z)
\overline{\Lambda}^{\mu\nu, kp+\tilde s}(\bz)$,  $k=0, 1, \cdots, 
\tilde{s}=\mu, \cdots, p-1$, disappear from the vertex operator 
$e^{h_{\mu\nu}\Phi(z,\bz)}$ because, as shown in 
eq.(\ref{eq:null}), $\Lambda^{\mu\nu}_{kp+\tilde{s}}$ 
are the elements in ${\cal X}_{\mu,\nu}$, 
the space of zero norm states, and so 
the vertex operator is given by means of the summation only of 
$\Lambda^{\mu\nu}_{kp+s}(z)\overline{\Lambda}^{\mu\nu,kp+s}(\bz)$, 
$s= 0,\cdots,\mu-1$, namely, the elements 
in the irreducible highest weight module 
$V^{irr}_{\mu,\nu}= V_{\mu,\nu}\backslash {\cal X}_{\mu_\nu}$. 
Therefore we can now use the important theorem in the Appendix, 
more concretely, the relation (\ref{eq:essential}): 
The irreducible highest weight module $V^{irr}_{\mu,\nu}$ 
is isomorphic to the tensor product 
$V_{\zeta}^{cl}\bigotimes{\cal V}_j $  
where $V_{\zeta}^{cl}$ and ${\cal V}_j$ are, respectively, 
the highest weight representation of 
the classical algebra $\slr$ with highest weight $\zeta$ 
and the $(2j+1)$-dimensional representation of $\qslc$. 
Owing to the isomorphism, the holomorphic sector factorizes  as 
\be
\Lambda^{\mu\nu}_{kp+s} (z)\mapsto \lambda^{\zeta}_k(z) 
\bigotimes \Psi^j_m(z),  \label{eq:iso}
\ee
and also does the anti-holomorphic sector.  
 
Keeping this remarkable  facts of $\ncom$ in mind, let us look at  the 
$N$-point correlation function in the Liouville theory 
associated with $\ncom$, 
\begin{eqnarray}
Z^S[{\sf m}:\mu_i,\nu_i] &:=& 
\langle V_{\mu_1\nu_1}(z_1,\bz_1)\cdots V_{\mu_N \nu_N}(z_N,\bz_N)
\rangle  \nonumber \\ 
{} &=& \int[d\Phi]e^{-\frac{1}{4\pi}\widehat{S}_L(\Phi;\hat{g})}
\prod_{i=1}^N V_{\mu_i \nu_i}(z_i,\bz_i), 
\label{eq:defcorr}  
\end{eqnarray}
where $V_{\mu_i \nu_i}(z_i,\bz_i) =e^{h_{\mu_i \nu_i}\gamma\Phi(z_i,\bz_i)}$,  
and ${\sf m}=(\sl{m},\overline{\sl m})$ 
stands for the moduli parameters of the surface. 
We will come back later to the discussions of correlation function and 
transition amplitude as well. 
Before that let us enjoy our model with the remarkable nature of $\ncom$. 
Thanks to the factorizations (\ref{eq:iso}) 
of the holomorphic sectors and the similar factorization 
of the anti-holomorphic sector,  
the vertex operators in (\ref{eq:defcorr}) are factorized as, 
\begin{eqnarray}
V_{\mu_i\nu_i}(z,\bz) &=& \sum_{k=0}^\infty N^{\zeta_i}_k 
\lambda^{\zeta_i}_k(z_i) 
\lambda^{\zeta_i,k}(\bz_i)\bigotimes\sum_{m=-j_i}^{j_i} {\cal N}^{j_i}_m 
\Psi^{j_i}_m(z_i)\Psi^{j_i,m}(\bz_i) \nonumber \\ 
 {} &=& e^{\zeta_i\cp(z_i,\bz_i)}\bigotimes 
e^{-j_i\tilde\gamma\phi(z_i,\bz_i)} 
\label{eq:separation}
\end{eqnarray}
where  $\cp(z,\bz)$ and $\phi(z,\bz)$ are, respectively, 
the classical and the quantum Liouville fields. 
With the remarkable decompositions of the vertex operators at hand, 
we can now  conjecture that the action also separates effectively as  
\be
\widehat{S}_L(\Phi;1\!\!\rm{I})\,\rightarrow\, 
S^{cl}(\cp;1\!\!\rm{I}) + S^q(\phi;\hat{g}_P).  
\label{eq:saction}
\ee
Here the actions $S^{cl}(\cp;1\!\!\rm{I})$ and $S^q(\phi;\hat{g}_P)$ 
are the Liouville actions with respect to the fields $\cf$ 
and $\phi(z,\bz)$, respectively, and are explicitly given by 
\begin{eqnarray}
S^{cl}(\cp;1\!\!\rm{I}) &=& \frac{1}{\beta^2}\int d^2z\left(
\partial_z\cp\partial_{\bz}\cp + \lambda  e^{\cp(z,\bz)}\right), 
\label{eq:caction}\\
S^q(\phi;\hat{g}_P) &=& \int d^2z 
\sqrt{\hat{g}_P}\left( 
\nabla\phi\overline{\nabla}\phi + QR_{\hat{g}}\phi(z,\bz) + 
\Lambda e^{\tilde\gamma\phi(z,\bz)}\right), 
\label{eq:waction}
\end{eqnarray}
where $\beta$ is some constant and 
$\hat{g}_P=e^{\tau\cp}dzd\bz$ is chosen as the background metric 
for  the quantum sector. 
When $\tau=1$ in the exponent of $\hat{g}_P$, the background 
metric is just the Poincar\'e metric. 
The reason why we have introduced new coupling constant 
$\beta$ is that, at this stage, we have no idea how the 
coupling constant of the classical sector should be.   
On the contrary, remembering that (see Appendix) the finite 
dimensional representations ${\cal V}_j$ appeared by the decomposition  
has the same deformation parameter $q$ as that of $V^{irr}_{\mu\nu}$,  
the renormalized coupling constant $\tilde\gamma$ of 
the quantum sector should satisfy, 
\be
\frac{\tilde\gamma^2}{2}\,=\, \frac{\gamma^2}{2}+2n, \,\quad\quad 
n=0,\,\pm1,\,\pm2,\cdots
\label{tgam}
\ee 
% However the relative strength $\kappa^2$ between the classical and 
% quantum sectors has been introduced and can be renormalized into 
% the coupling constant $\gamma$, 
% $\gamma\rightarrow \tilde\gamma+\gamma/\kappa$. 
It should be emphasized that, upon the condition 
$\gamma Q_0=\tilde\gamma Q=1$, the conjecture (\ref{eq:saction}) 
together with (\ref{eq:caction}),(\ref{eq:waction}) 
is exactly true under the substitution 
$\gamma\Phi=\tau \cp +\tilde\gamma\phi$ with $\tau=\gamma/\beta$.  

Some remarks are now in order. 
The above observations seems to suggest that, in the 
quantum Liouville theory via $\ncom$, the Liouville field 
$\Phi(z,\bz)$  should be expanded as 
$\gamma\Phi=\tau\cp+\tilde\gamma\phi$ and,  
therefore, one can  interpret the field $\phi$ as the quantum fluctuation 
around $\tau\cp$. 
In other words, the metric $ds^2=e^{\gamma\Phi(z,\bz)} dz d\bz$ 
is to be written as $ds^2=e^{\tilde\gamma\phi(z,\bz)}{\hat g}_{P}$  
and $e^{\tilde\gamma\phi(z,\bz)}$ represents quantum fluctuation of metrics 
around the classical background metric ${\hat g}_{P}=e^{\tau\cp} dz d\bz$.  
If $\tau$ is set to be 1, the background metric ${\hat g}_{P}$ 
becomes the Poincar\'e metric. 
Now we can give a possible interpretation of quantum $2D$ manifold     
according to the above observations: 
{\em Quantum manifold  with metric $e^{\gamma\Phi(z,\bz)}$ should be 
considered as the total system of 
$2D$ classical manifold  with the metric $e^{\tau\cp} dz d\bz$ 
and quantum fluctuations around the classical surface}. 
This interpretation matches quite well to the general concept 
of quantum object.   
% From the viewpoint of eq.(\ref{eq:start}), 
% the action $S^q(\phi;\hat{g}_P)$ in (\ref{eq:waction}) has arisen in the way 
% $Z[e^{\kappa\phi} {\hat g}_P]=\exp (-const\cdot S^q(\phi))Z[{\hat g}_P]$. 
Note also that the action $S^q(\phi)$ governing the quantum 
fluctuations is again the quantum Liouville theory 
associated with $\com$. 

It is worthwhile to comment here the difference between our 
formulation and the standard weak coupling Liouville theory 
\cite{Po2}--\cite{DK}.  
In the standard approaches 
whose action is given in (\ref{waction}),  
the theory also has $\com$ structure as shown in section 3.1. 
But unlike the quantum sector $S^q(\phi;\hat{g}_P)$ of our model, 
the background metric can be freely chosen together with 
the shift of Liouville field $\Phi(z,\bz)$.  
For example, in Refs.\cite{Po2,KPZ},  
the metric in the light-cone gauge $ds^2=\vert dx_+ + \mu dx_-\vert^2$,  
which is conformally equivalent to the metric in the conformal gauge 
$ds^2=e^{\tilde\gamma\tilde\phi(z,\bz)}dz d\bz$ 
was chosen. 
These choices have the flat background metric. 
On the contrary, in our formulation, the background metric is 
automatically selected as $\hat{g}_P$. 
Thus the appearance of the classical sector in addition to the 
quantum theory associated with $\com$ is the characteristic 
feature of our formulation. 
We will recognize  the importance of the classical sector 
from the viewpoints of geometric quantization of moduli space 
in section 3.5.

\subsection{Fusion Rules}

Before going to the discussion of the correlation function, 
it is interesting to look at fusion rules in our model. 
First, we have to investigate Clebsh-Gordan (CG) decomposition rule 
for the tensor product of two irreducible infinite dimensional 
representations of $\qslc$,
\begin{equation}
V_1 \otimes V_2\;\longrightarrow\; V_3,
\end{equation}
where $V_i:=V_{\mu_i, \nu_i}^{irr}\cong V_{\zeta_i}^{cl}\otimes 
{\cal V}_{j_i}$. 
The quantum CG coefficient, known as the $q$-$3j$ symbol, 
for the infinite dimensional representations of $\qslc$ 
is given when $q$ is $not$ a root of unity as follows,  
\begin{eqnarray}
&&\left[ \begin{array}{ccc} h_1 & h_2 & h_3 \\
                             r_1 & r_2 & r_3 \end{array}\right]^{\qslc}_q
=C(q)\delta_{r_1+r_2,\,r_3}\tilde{\Delta}(h_1, h_2, h_3)  \nonumber \\
&&\times \left\{
\frac{[2j-1][r_3-h_3]![r_1-h_1]![r_2-h_2]![r_1+h_1-1]![r_2+h_2-1]!}
      {[r_3+h_3-1]!}\right\}^{1/2} \nonumber \\
&&\times\sum_{R\ge 0}(-)^R q^{\frac{R}{2}(r_3+h_3-1)}
                \frac{1}{[R]![r_3-h_3-R]![r_1-h_1-R]![r_1+h_1-R-1]!}
\nonumber\\
&&\quad\cdot\frac{1}{[h_3-h_1-r_1+R]![h_3+h_2-r_1+R-1]!},
\label{CG}
\end{eqnarray}
where $C(q)$ is a factor which is not important for our analysis below and
\begin{eqnarray}
&{}&\tilde{\Delta}(h_1, h_2, h_3)  \\
&{}&\quad =\{[h_3-h_1-h_2]![h_3-h_1+h_2-1]!
                        [h_3+h_1-h_2-1]![h_1+h_2+h_3-2]!\}^{1/2}. \nonumber
\end{eqnarray}
The notations $h_i:=h_{\mu_i\,\nu_i}=\zeta_i p-j_i, \;
r_i=(\zeta_i+k_i)p-m_i$ have been used.
Of course for our case, \ie, $q=\exp(\pi i \frac{p'}{p})$, 
the CG coefficient is not necessarily well-defined 
due to the factor $[p]=0$.
What we have to do is only to find conditions which give  
finite CG coefficients. 
Since the calculation is lengthy and our interest is not in the details, 
we describe here only the quite interesting result;  
finite CG coefficients exist if and only if
\begin{equation}
\begin{array}{l}
\zeta_1 + \zeta_2 \le \zeta_3, \\
{}\\
\vert j_1 - j_2 \vert -1 < j_3 \le {\rm min}\,(j_1+j_2, p-2-j_1-j_2).
\end{array}
\end{equation}
It should be noticed that, on the modules $V^{cl}$ and ${\cal V}$, 
the coproducts of the 
operators $K=q^H$ and $L_0$ are $\Delta(K)=K\otimes K$ and 
$\Delta(L_0)=L_0\otimes 1 +1\otimes L_0$, respectively. 
The coproduct of the Cartan operator yields the conservation
law of the highest weights, physically speaking, the conservation of the
spin or angular-momentum along the $z$-axis.
Now, from the above coproducts, we have the conservation laws
$(\zeta_1+k_1) + (\zeta_2+k_2) = (\zeta_3+k_3)$ and $m_1 + m_2 = m_3$
(mod $p$). 
Therefore the minimum value of $j_3$ is just $\vert j_1-j_2 \vert$ because
the difference between $\vert j_1-j_2 \vert$ and $j_3$ is always integer.
We therefore obtain the following decomposition rule of the tensor product of
two infinite dimensional representations of $\qslc$ at a root of unity,
\begin{equation}
\left( V_{\zeta_1}^{cl}\otimes {\cal V}_{j_1}\right) \bigotimes
\left( V_{\zeta_2}^{cl}\otimes {\cal V}_{j_2}\right)
= \left( \bigoplus_{\zeta_1+\zeta_2\le\zeta_3}V_{\zeta_3}^{cl}\right)
\bigotimes \left(\bigoplus_{j_3=\vert j_1-j_2\vert}^{{\rm min}\,
\{j_1+j_2, p-2-J_1-j_2\}}{\cal V}_{j_3}\right).
\end{equation}
The decomposition rules for the tensor products of $V_\zeta^{cl}$ 
and of ${\cal V}_j$ are the same as those for the tensor products 
of the $\slr$ and of the finite dimensional 
representation of $\qslc$, respectively. 

Applying this decomposition rule to our model, 
fusion rule is 
\be
\left[e^{h_1\gamma\Phi}\right]\times \left[e^{h_2\gamma\Phi}\right]
=\left(\sum_{\zeta_1+\zeta_2\le\zeta_3}\left[e^{\zeta_3\cp}\right]
\right)\bigotimes \left(
\sum_{j_3=\vert j_1-j_2\vert}^{{\rm min}\,
\{j_1+j_2, p-2-J_1-j_2\}}\left[e^{-j_3\tilde\gamma\phi}\right)\right), 
\ee
namely, exactly the same as the tensor product the fusion rules 
of the classical theory and those of the weak coupling theory. 
This result means that the classical sector and the quantum sector never 
mix with each other. 

\subsection{Correlation Functions and Amplitudes}

Let us return to the discussion of the correlation functions 
in our model. 
The original definition was given in eq.(\ref{eq:defcorr}). 
What we will see in this section is how the correlation function 
and amplitude can be written by the decompositions of vertices and actions. 
Since, as we have seen, the Liouville field $\Phi$ can be expanded around 
the classical solution, the functional measure is  
$[d\Phi]=[d\phi]$ up to a constant.  
The correlation function (\ref{eq:defcorr}) is expected finally 
to be factorized into classical sector and quantum sector, 
\ie, up to some constant, it can be written as 
\be
Z^S[{\sf m}:\{\mu,\nu\}] = 
Z^{cl}[{\sf m}:\{\zeta\}]Z^q[{\sf m}:\{j\}].    
\label{eq:correlation}
\ee
Here the quantum sector is given by 
\be
Z^q[{\sf m}:\{j\}]:=\int [d\phi] 
\,e^{-\frac{1}{4\pi} S^q(\phi;\hat{g}_P)}\,
\prod_{i=1}^N  e^{-j_i\phi(z_i,\bz_i)},  
\label{eq:qpartition}
\ee
and the classical sector is  
\be
Z^{cl}[{\sf m}:\{\zeta\}]=e^{-\frac{c_0}{48\pi}\overline{S}^{cl}(\cp)},
\label{eq:classical}
\ee
where $c_0=12/\beta^2$ is the central charge of 
the classical Liouville theory. 
The action $\overline{S}^{cl}(\cp)$ 
denotes the classical action defined on the surface with 
$N$ branch points and  regularized by subtracting the singularities 
near the branch points. 
In particular, in the case when the topology is the 
$N$-punctured sphere,  \ie, 
all the vertices carry the charges 
$\zeta_i=1/2\beta^2$, $\overline{S}^{cl}(\cp)$ is 
given by eq.(\ref{eq:Action}). 
Let us discuss in more detail.  
As stated above, the action $S^q(\phi;\hat{g}_P)$ corresponds to 
the Liouville action in the conformal gauge 
$ds^2=e^{\tilde\gamma\phi(z,\bz)}{({\hat g}_P)}_{z\bz}dz d\bz$ with 
$\hat{g}_P$ being the classical background. 
% The central charge coming from this sector is 
% \be
% c_q=1+12Q^2.   \label{eq:weakcenter}
% \ee
Suppose that the correlation function $Z^q[{\sl m},\{j\}]$ 
admits holomorphic factorization as, 
\be
Z^q[{\sl m}:\{j\}] =\sum_{I,J}N^{I,J}\overline{\Psi}_I[\overline{m}:\{j\}]
\Psi_J[m:\{j\}],  \label{eq:holomorphic}
\ee
where $N^{IJ}$ is some constant matrix. 
Here $\Psi_I[m:\{j\}]$ is expected to be a holomorphic section of 
a line bundle over the moduli space ${\cal M}_{0,N}$. 
We will see later that this expectation is acceptable. 

Now we are at the stage to discuss transition amplitudes 
of our Liouville theory. 
In order to obtain a transition amplitude, we have to integrate 
the correlation function $Z[{\sf m}]$ over the moduli space of surface 
metric with the Weil-Peterson metric. 
Combining (\ref{eq:classical}) and (\ref{eq:holomorphic}) 
and integrating over the moduli space, $N$-point transition amplitude 
${\cal A}_N(\{\zeta\},\{j\})$ has the following form,   
\be
{\cal A}_N(\{\zeta\},\{j\}) = \sum_{I,J}N^{I,J}
\int_{{\cal M}_{0,N}} d(WP) e^{-\frac{c_0}{48\pi}\overline{S}^{cl}}
\overline{\Psi}_I[\overline{m};\{j\}]\,\Psi_J[m:\{j\}], 
\label{eq:amplitude}
\ee
where $d(WP)$ stands for the Weil-Peterson measure on the moduli space. 
Here let us see the geometrical meaning of the classical sector. 
For the topology of the $N$-punctured sphere, 
there are quite remarkable facts shown by Zograf and Takhtajan \cite{ZT}  
about the connection between the classical Liouville theory 
and K\"ahler geometry of the moduli space of complex structure. 
The important facts are as follows: 
First the Liouville action evaluated on the classical solution is 
just the K\"ahler potential of the Weil-Peterson symplectic structure, 
precisely, $\omega_{WP}=i\overline{\partial}\partial \overline{S}^{cl}/2$.  
Second the accessory parameters $c_i, i=1\sim N$ are written as 
$-2\pi c_i=\partial \overline{S}^{cl}/\partial z_i$.\footnote{This 
relation between the classical Liouville action and the accessory
parameter which is associated with every puncture was first 
conjectured by Polyakov.\cite{Po3}} 
>From these facts, one can show that the functions $c_i$ are in 
involution \cite{Ta3}, \ie, $\{c_i, z_j\}_{WP}=i\delta_{ij}$, 
where $\{\,\,,\,\,\}_{WP}$ 
is the Poisson bracket with respect to the Weil-Peterson symplectic 
2-form $\omega_{WP}$. 
Thus the classical Liouville theory can be regarded as the 
K\"ahler geometry of the moduli space of surfaces. 

Due to the relation, it is natural to expect the deep connection 
between our quantum Liouville theory and the geometric quantization 
of the moduli space which corresponds to 
the phase space of the classical geometry. 
With this hope, we should try to observe our result 
from the viewpoints of geometric quantization. 
Before doing this, it is helpful to summarize the basic 
facts about the geometric quantization of a classical theory.  
Consider a K\"ahler manifold ${\cal M}$ equipped with a symplectic 
structure $\omega$ which is written in terms of the K\"ahler
potential $K$ as $\omega=i\partial\overline{\partial} K$.  
The K\"ahler manifold plays the role of the phase space 
of the classical theory. 
Geometric quantization is performed by building a line bundle 
${\cal L}\rightarrow {\cal M}$ over the $2N$ dimensional 
manifold ${\cal M}$ with the curvature two form $F= -i\omega$. 
Let us parameterize ${\cal M}$ by $(q_i, p_i),\, i=1\cdots N$, 
and denote the section as $\psi(q_i, p_i)$. 
The final manipulation to complete the geometric quantization 
is to impose on ${\cal L}$ the condition that 
the section is annihilated by the derivatives of 
half of the variables, which is called a polarization.  
In other words, by the  choice of a polarization, 
sections are represented by $N$ variables.  
Let us, for example, choose the polarization as 
$\partial_{p_i}\psi=0,\, i=0,\cdots,N$. 
Then the Hilbert space ${\cal H}$ is the space of sections with the 
polarization, 
\be
{\cal H}=\{\psi\,\vert\, \partial_{p_i}\psi=0\}
\ee
The inner product on ${\cal H}$ is
introduced with the Hermitian metric $e^{-K}$, \ie, 
for $\psi_1, \psi_2 \in {\cal H}$, it is  
$\langle\psi_1, \psi_2\rangle = \int 
e^{-K}\overline{\psi}_1\cdot \psi_2$,   
where the measure is that defined by $\omega$. 
Putting our expression (\ref{eq:amplitude}) together with 
the fact that, as we have seen in the last part of Section 2, 
a line bundle with the Hermitian metric 
$\exp(\frac{k}{\pi}\overline{S}^{cl})$ 
can be constructed on the moduli space 
${\cal M}_{0,N}=({\sl m}_i, c_i)$,  
the holomorphic part $\Psi_I[m]$ of the quantum sector can be 
regarded as a holomorphic section of the  
Hermitian line bundle ${\cal L}_{c_0}\rightarrow {\cal M}_{0,N}$ 
with the polarization $\partial_{c_i}\Psi=0$ and 
the curvature is $\frac{c_0}{12\pi}\omega_{WP}$. 
On the other hand, the classical correlation function 
(\ref{eq:classical}) corresponds to a Hermitian metric 
defining an inner product $\langle\,\,,\,\,\rangle_{c_0}$  
on the Hilbert space which is the space of sections $\Psi[m]$. 
Hence, at least for the topology of the $N$-punctured sphere, 
the amplitude can be written as 
${\cal A}=\sum N^{I,J}\langle \Psi_I , \Psi_J\rangle_{c_0}$. 

Thus the quantum Liouville theory associated with $\ncom$ 
fits well with the geometric quantization of moduli space. 
The factorization property into the classical sector and 
the quantum sector plays an important role here and is just the special 
feature appearing only in the Liouville theory associated with
$\ncom$.

\subsection{Central Charge and Some Discussions}

Finally we give some discussions about our Liouville theory.   
What we have understood in the above discussions is that 
the quantum Liouville theory based on the infinite dimensional 
representations of $\qslc$ at a root of unity factorizes 
into the classical Liouville theory and  the quantum Liouille theory 
based on the finite dimensional reprsentations of $\qslc$, 
and the latter governs quantum fluctuations around the 
classical surface.  
% Not only vertex operators but also fusion rules 
% can be considered as the tensor products 
% of those in the classical and weak coupling theories.  
The following discussions are on this observation, namely, 
we start from the action 
\be
\frac{1}{4\pi}\widehat{S}_L=\frac{1}{4\pi\beta^2}\int d^2z \left( \partial \cp 
\overline{\partial}\cp + e^{\cp}\right) +
\frac{1}{4\pi}\int d^2z \sqrt{\hat{g_P}} \left(\nabla \vp
\overline{\nabla}\vp+ Q R_{\hat{g_P}}\vp \right), 
\label{totaction}
\ee
rather than the original one (\ref{eq:qaction}). 
Here we have set $\Lambda=0$ for the convenience. 
It is important to notice that,  since  $e^{\gamma\Phi}$ is a 
Riemannian metric, the Liouville field $\gamma\Phi$ should be real,  
and, therefore, $\tilde\gamma \phi$ is real as well. 
Recall the relation (\ref{tgam}) and notice that $\tilde\gamma^2$ 
can be negative. 
When $\tilde\gamma$ is pure imaginary, the quantum filed $\vp$ is 
also imaginary and the second term in (\ref{totaction}) has 
wrong sign, while if $\tilde\gamma$ is real, $\vp$ is a real field 
and the quantum sector has correct sign. 
In the following discussions, 
we would like to impose another assumption that 
$\beta=\gamma$, equivalently, $\tau=1$, in order for the background 
metric $\hat{g}_P$ to be the poincar\'e metric, 
  
It is easy to calculate EM tensor from the total action, 
and one finds, 
\begin{eqnarray}
T^{tot}_{zz}&=&T^{cl}(z) + T^q(z) + T^{\rm{mix}}(z), \label{emtot} \\ 
&& T^{cl}(z)=\frac{1}{\gamma^2}\left(-\frac{1}{2}\partial \cp \partial 
\cp+ \partial^2\cp\right),  \nonumber \\[.1cm]
&&T^q(z)=\lim_{w\rightarrow z}\left[\left(-\frac{1}{2}
\partial_z \vp \partial_w \vp + Q\,\partial^2 \vp\right)
-\frac{1}{(z-w)^2}\right],\nonumber \\[.1cm] 
&&T^{\rm{mix}}(z)=Q\partial \cp \partial \vp, \nonumber \\[.1cm]
T^{tot}_{z\bz}&=& 0. \label{totEM}  
\end{eqnarray}
The term $T^{\rm{mix}}$ comes from the curvature term 
in (\ref{totaction}) where the classical and quantum 
Liouville fields $\cp,\, \vp$ interact with each other. 
The second line (\ref{totEM}) guarantees that the total system is 
again conformally invariant. 

It is the time to estimate the central charge to confirm thet 
our model is actually a  strong coupling theory. 
To do this, let us observe how the EM tensor $T(z)\equiv T^{tot}_{zz}$ 
transfroms under the change of variable $z\rightarrow f(z)$. 
Since the EM tensor is a second rank tensor with central extension, 
it should satisfy the following transformation law, 
\be 
\hat{T}(z)=\left(\frac{df}{dz}\right)^2T(f)+\frac{c_L}{12}
\{f,z\}_S. \label{EMlaw}
\ee
However, one sees that eq.(\ref{emtot}) satisfies (\ref{EMlaw}) 
if and only if $Q=0$, and the central harge is given by 
\be
c_L=1+\frac{12}{\gamma^2}. \label{cofstrong}
\ee
The first term 1 arises from the quantum sector $T^q(z)$, precisely, 
from the subtraction of the singularity $1/(z-w)^2$, \footnote{
The author thanks N. Ano for a very useful discussion on this}   
and the second term $12/\gamma^2$ comes from the classical sector. 
Puttig together the central charge (\ref{cofstrong}) with eq.(3.8), 
one finds  
\be
\frac{1}{\gamma} = \sqrt{\frac{25-D}{12}}=Q_0. 
\label{gam}
\ee
Notice that the original coupling constant $\gamma$ is real 
for the dimension $D\leq25$, and that 
the central charge of our Liouville theory 
can be actually in the strong coupling region, 
$1<c_L<25$ for $\gamma>1/\sqrt{2}$.  
It is interesting to give some of the allowed 
dimension $D$ of the target space in our model. 
Recalling the assignment $\gamma^2/2=p'/p$ with integers 
$p,p'$ being coprime with each other and (\ref{gam}), 
one finds \\[.05cm]
\begin{eqnarray}
&({\rm i})&  \mbox{when}\,\,p'=1, \quad D=1, 7, 13, 19, 
\nonumber \\[.1cm]
&({\rm ii})&  \mbox{when}\,\,p'=2, \quad D=4, 10, 16, 22, 
\label{dim} \\[.1cm]
&({\rm iii})&  \mbox{when}\,\,p'=3, \quad D=3,5,9,15,17,21,23. 
\nonumber \\
&&\quad\quad\quad \cdots\cdots  \nonumber
\end{eqnarray}
We can obtain fractional dimensions as well. 
If another matter whose central charge is fractional couples 
to the string theory, such a fractional 
dimension will become crucial. 
In the case (i), the dimensions are completely the same as those given 
in \cite{Gerv}. 
This is a strange coincidence, since we are dealing with 
the case when $q$ is a root of unity, whereas, in \cite{Gerv}, 
the parameter $q$ is not a root of unity. 

This central charge (\ref{cofstrong}) 
coincides with that obtained by Takhtajan in 
Refs.\cite{Ta,Ta2}, where a manifestly geometrical approach 
was used. 
Along the Polyakov's original formulation of quantum Liouville 
theory, he started with conformal Ward identities via functional 
integral. 
Upon perturbation expansions around the classical solution 
corresponding to the Poincar\'e metric, he obtained 
the central charge (\ref{cofstrong}). 
It is worth mentioning that in both approaches, 
the quantum Liouville theory is expanded around the classical solution, 
\ie, the Poincar\'e metric. 
In other words, a quantum Riemann surface is treated  as a 
total of a classical surface endowed with the 
hyperbolic geometry and quantum corrections around it. 
Indeed, in Takhtajan's approach, the classical limit recovers 
the underlying hyperbolic geometry of Riemann surface. 
In our case, owing to the condition $Q=0$,  
the total action (\ref{totaction}) is just the sum of the 
clasical Liouville action and a quantum action 
without any mixing terms.   
The quabtu field $\vp(z,\bz)$ can be interpreted as the 
$D+1$-th component of the string, and we obtain 
the $D+1$ dimnsional spacetime.  
As suggested previously, the signature the spacetime has 
depends on the sign of $\tilde\gamma^2$.  
  
Next we turn our attention to the conformal dimension. 
In also this discussion, we are on the conjecture $Q=0$. 
The vertex operator of the string dressed by gravity is written as 
\be
V(z,\bz)=\int d^2z \sqrt{g}:{\cal O}_X :\quad \cong \quad 
\int d^2z e^{\tau\cp}: e^{\tilde\gamma\vp}{\cal O}_X: 
\ee
where ${\cal O}_X$ stands for a pure string vertex operator. 
Let $\Delta_0, \Delta_X$ be the dimensions of the classical Liouville 
exponential  and the string vertex, respectively, and 
one finds the following relation to the conformal dimensions
\be
\Delta_0-\frac{\tilde\gamma^2}{2}+\Delta_X=1. 
\label{relation1}
\ee
The middle term in the left hand side of (\ref{relation1}) is the  
conformal dimension of the exponential $e^{\tilde\gamma\vp}$. 
Since we have chosen  $\tau=1$, 
the background metric $\hat{g}_P$ is  the 
Poincar\'e metric and so $\Delta_0=1$.  
Then, together with (\ref{tgam}), one finds  
\be
\frac{\tilde\gamma^2}{2}=\Delta_X=\frac{\gamma^2}{2}+2n, 
\qquad n=0,\,\pm1,\,\cdots.
\label{Xdim}
\ee
This equation suggests that the matters (string) which couple  
to the Liouville theory form a discrete series in conformal dimension 
and that the constant $\tilde\gamma$ can be imaginary.

In summary, our Liouville theory can be 
interpreted as follows: 
It is the total system of the classical Liouville theory plus the 
fluctuation around the classical manifold, 
which becomes the  $D+1$-th component of the string. 
The same situation occurs in the weak coupling Liouville theory, 
although in that case the classical Liouville sector does not appear.  
In the weak coupling theory when the dimension of the target space 
is $D=25$, the quantum Liouville field is 
regarded as the time component of the string with wrong sign 
and,  as the result, the spacetime becomes the 
26 dimensional Minkowski space. 
On the contrary, in our case, we have more choices for the 
dimension $D$ as listed in (\ref{dim}), especially,  
for $\gamma^2=\frac{6}{11}$, the dimension of the 
target space is $D=3$ and, therefore,  with the field $\vp$ 
as the time component, our space-time is the $4$ dimensional 
Minkowski or Euclid space according to the sign of the quantum 
sector $S^q$.

%%%%%%%%%%%%%%%%%%%%%%%%%%%%%%%%%%%%%%%%%%%%%%%%%%%%%%%%%%%%%%%%%%
\section{Summary and Discussions}

We have developed a new  approach to the quantum Liouville theory  
via the infinite dimensional representations 
of $\qslc$ when $q=\exp (\pi i\, p'/ p)$. 
In this Liouville theory we have dealt with, 
only the vertex operators with positive charges 
$\alpha$ can be defined. 
The characteristic feature of $\ncom$ is that every irreducible 
highest weight module $V^{irr}_{\mu,\nu}$ necessarily factorizes 
into the highest weight module $V_\zeta^{cl}$ of the classical 
algebra $\slr$ and ${\cal V}_j$, the $(2j+1)$-dimensional 
representation of $\qslc$. 
Our investigations in this article have been performed by making full 
use of this feature. 
Owing to this fact, we observed that the vertex operators with positive 
charge $\alpha=h_{\mu\nu}$ factorized into the {\em classical} vertex 
operator with charge $\zeta$ and the vertex operator with negative 
charge $-j$ as in eq.(\ref{eq:separation}), 
where the relation $h_{\mu\nu}=\zeta p -j$ was understood. 
We further conjectured that the Liouville action 
$\widehat{S}_L(\Phi)$ should also 
decompose into the classical Liouville action $S^{cl}(\cp)$ and 
$S^q(\phi;\hat{g}_P)$ with the classical background 
$\hat{g}_P=e^{\tau\cf}$, where $e^{\cp}$ defines the Poincar\'e metric. 
Since the field  $\phi(z,\bz)$ can be interpreted as the Liouville 
field which measures the quantum fluctuations of the metric 
around the classical metric $e^{\tau\cp}$, 
the Liouville theory $S^q(\phi;\hat{g}_P)$ governs the theory of 
quantum fluctuations of quantum Riemann surfaces around 
the classical surfaces. 
Namely in our formulation, the quantum Liouville theory 
$\widehat{S}_L(\Phi)$ 
describes quantum $2D$ space-time as the total system 
of the classical space-time and 
the quantum fluctuations around it. 
We found that the Liouville theory governed by the action 
$S^q(\phi;\hat{g}_P)$ was associated with the {\em finite} dimensional 
representations $\com$. 
We remarked the difference between our Liouville theory 
and the standard Liouville theory  \cite{Po2}--\cite{DK}  
which is confined to the weak coupling region. 
Although, in both theories, quantum fluctuations of metric are 
deeply related to $\com$,  
our theory contains the classical Liouville theory yielding 
the underlying K\"ahler geometry, whereas  the latter does not. 
It is quite important to emphasize that,  
our  quantum Liouville theory is certainly in the strong coupling regime.  
% and in that case the quantum 
% fluctuation around the classical space-time yields matter and, 
% therefore,  $S_L(\Phi)$ can be interpreted as the unified action 
% of matter and classical gravity. 
 
Another interesting result was that our Liouville theory  
fitted well with the concepts of geometric quantization of the 
moduli space of metrics. 
First of all, we have noted that the classical Liouville theory is 
in agreement with the K\"ahler geometry of the moduli space, where 
the Liouville action evaluated on the classical solution $\cp$  
is nothing but the K\"ahler potential of the Weil-Peterson metric 
$\omega_{WP}$. 
Second, we have observed that 
the Hermitian line bundle ${\cal L}_{c_0}$ can be built 
over the moduli space ${\cal M}_{0,N}$ with the Hermitian metric 
$\exp\,(-\frac{c_0}{48\pi}\overline{S}^{l})$. 
The role of holomorphic sections of the line bundle ${\cal L}_{c_0}$ 
is played by the holomorphic part $\Psi[m;\{j\}]$ of the quantum 
sector $Z^q[{\sl m}:\{j\}]$. 
Thus, as we have seen in eq.(\ref{eq:amplitude}), 
the transition amplitude in our theory agrees  with 
the inner product of two wave functions corresponding to the 
sections of the line bundle ${\cal L}_{c_0}$. 
Remembering that the quantum Liouville theory can be regarded as 
a quantum geometry of Riemann surfaces, our observation of our quantum 
Liouville  theory from the viewpoints of geometric quantization of 
moduli space seems to be quite natural. 

At this stage, it is worthwhile to compare our quantum gravity with  
that formulated in Ref.\cite{Ver} where $c>1$ two-dimensional 
quantum gravity was treated via geometric quantization of 
moduli space. 
Riemann surfaces in his consideration are compact and have 
genus $g>1$. 
There, the transition amplitude between some initial state 
$\Psi_I$ and some final state $\Psi_F$ is given by the inner product 
(see eq.(5.7) in Ref.\cite{Ver})  
\be
\langle \Psi_I, \Psi_F\rangle =\int d(WP) Z_L[{\sl m}] 
\overline{\Psi}_I[\overline{\sl m}] \Psi_F[{\sl m}], \label{eq:Verlinde}
\ee
where $Z_L[{\sl m}]$ 
is the Liouville partition function (let $\sigma(z,\bz)$ be the 
Liouville field). 
$\Psi[m]$ is a holomorphic section of a line bundle over the moduli 
space and can be identified with the conformal block obtained by 
solving the conformal Ward identity of Polyakov \cite{Po2}. 
Therefore $\Psi[m]$ is considered as the holomorphic sector 
of the partition function in the weakly coupled Liouville gravity. 
In that paper, the metric on the Riemann surface is parameterized as 
$ds^2=e^{\sigma(z,\bz)}\vert dz + \mu d\bz\vert^2$. 
Now it is easy to find intimate relations  between the formulation 
of Verlinde and our Liouville gravity except only one big difference. 
In both theories, a quantum Riemann surface is composed of  two 
sectors, one is the Riemann surfaces as a background and the other 
corresponds to quantum fluctuations around the background surface. 
In the Verlinde's formulation, the quantum fluctuation is 
parameterized by the Beltrami differential $\mu$. 
The difference arises in the choice of background surfaces; 
the background Riemann surface in our theory is just 
the classical surface with the Poincar\'e metric $e^{\cf}$, 
while in the Verlinde's formulation it is again a quantum surface 
with the metric $e^{\sigma(z,\bz)}$. 
Because $\sigma(z,\bz)$ is not a classical field, the partition 
function $Z_L[{\sl m}]$ in eq.(\ref{eq:Verlinde}) 
cannot be written as $Z^{cl}[{\sl m}]$ in our theory. 
Note again that the holomorphic sections $\Psi[m]$ in both theories 
are related only to the quantum fluctuations. 

\vspace{.2cm}

Although we have observed some remarkable features of the quantum 
Liouville theory associated with $\ncom$ and obtained a natural 
concept of quantum $2D$ space-time within the framework of our 
formulation, 
there still remain some important problems to be investigated.  
I list some of them below. 
First, and maybe most important problem is the explicit 
relation between our quantum Liouville theory  
and Takhtajan's one. 
These two models give the same central charge.   
Moreover, in both approaches, the classical geometry 
appears as a background, namely, the quantum metrics 
are to be expanded around the classical metric of Riemann surface, 
the Poincar\'e metric.  
Inspired by the agreements, it is quite interesting and 
important to find the relation, although at first glance 
these two approaches are completely different, \ie, Takhtajan's approach 
is fully  geometric and ours is algebraic.   
Second, Virasoro structure of this model is an interesting problem. 
Finding this structure will allow more explicit discussions 
of our Liouville theory, especially correlation functions. 
This also maybe sheds light on the problem of $c>1$ discrete 
series of the Virasoro algebra. 
It is well-known that representations of the Virasoro algebra 
with $c>1$ always form continuous series. 
However, since the highest weight representations $\ncom$ form a 
discrete series and our model can have central charges greater than 1, 
it can be expected to find  such a discrete series. 
These discussions will appear elsewhere.\cite{AS}

\vspace{.5cm}
\noindent 
{\bf Acknowledgment :}\,\, I wish to thank  Profs. L. D. Faddeev and 
L. Takhtajan for communications on strongly coupled Liouville gravity. 
It is also a pleasure to acknowledge Profs. M. Ninomiya,  R. Sasaki 
and Dr. N.Ano for discussions and comments.

%%%%%%%%%%%%%%%%%%%%%%%%%%%%%%  APPENDIX  %%%%%%%%%%%%%%%%%%%%%%%%%%%%%%%

\appendix 
\section{Infinite Dimensional Representations with $q$ at a root of unity} 

The Appendix gives a brief review of Ref.\cite{MS} 
where the infinite dimensional highest weight representations of 
$\qslc$ is examined for the case 
when the deformation parameter $q$ is a root of unity.  
Let $q=\exp\,\pi i/p$. \footnote{In the previous sections, 
we used the choice $q=\exp\,\pi i \frac{p'}{p}$.  
For the brevity of our discussions, we will choose $p'=1$ in the 
appendix.  
The essential parts of our discussion is independent of 
the choice for $p'$. }
The most essential feature which we have made full use of in our 
discussions is stated in the following theorem; 

\vspace{.3cm}
\noindent
{\bf Theorem}. \hspace{.3cm} 
Every {\em infinite} dimensional irreducible highest weight
representation, denoted as $V^{\qslc}_I$ 
is necessarily isomorphic to the tensor product 
of two highest weight representations as  
\be 
V^{\qslc}_I \cong V^{\slr} \bigotimes V^{\qslc}_F, 
\label{eq:theorem}
\ee
where $V^{\slr}$ is a representation of the {\it classical} algebra 
$su(1,1)\cong\slr$ and $V^{\qslc}_F$ is a {\em finite} dimensional 
representation of $\qslc$. 

\vspace{.3cm}
\noindent
Below we will prove this theorem (see Refs.\cite{MS} for the detailed
discussions). 

Let $X_+, \, X_-, K=q^H$ be the  generators of the quantum universal 
enveloping algebra $\qslc$ satisfying the relations 

\be
[X_+, X_-]=\frac{K^2-K^{-2}}{q-q^{-1}}, 
\quad K X_{\pm}=q^{\pm 1}X_\pm K. 
\label{eq:qslcrelation}
\ee
To get infinite dimensional representations, 
we define hermitian conjugations as  
$X^\dagger_\pm = -X_\mp, \, K^\dagger = K^{-1}$. 
As in the classical case, we represent $\qslc$ with these 
conjugations by constructing the highest 
weight module $V_h$ on the highest weight state $\Lambda^h_0$ such as 
$X_-\Lambda^h_0 =0, \,  K\Lambda^h_0=q^h\Lambda^h_0$,  
\be
V_h=\{\Lambda^h_r\, \vert \,\Lambda^h_r :=\frac{X_+^r}{[r]!}
\Lambda^h_0, \,\, r=0, 1,\cdots \,\}. 
\ee
Upon using the hermitian conjugations and the relations 
(\ref{eq:qslcrelation}), 
the norm of the state $\Lambda^h_r$ is easily calculated as  
\be
\parallel \Lambda^h_r \parallel^2 = \left[ \begin{array}{c} 
                                         2h + r - 1 \\ r 
                                  \end{array} \right]_q
\label{eq:norm}
\ee 
with the normalization $\parallel \Lambda^h_0 \parallel^2=1$. 

The first problem we come across when $q$ is a $p$-th root of unity is
that the norm of the state $\Lambda^h_p$ diverges owing to the factor 
$[p]=0$ in the denominator. 
Thus, for arbitrary highest weight $h$, 
the highest weight module $V_h$ is not necessarily well-defined. 
The only way to avoid this undesirable situation is to require that   
there exist two integers $\mu\in\{0, 1, \cdots, p\}$ and 
$\nu\in {\N}$, such that the highest weight is given by, 
\be 
h=h_{\mu\nu}:=\half(p\nu -\mu +1).    
\label{eq:highest}
\ee
For the highest weight $h_{\mu\nu}$, 
the factor  $[2h_{\mu\nu}+\mu-1]=0$ appearing in the numerator of the 
right hand side of the norm (\ref{eq:norm}) makes 
the $p$-th state a finite-norm state, and so the module 
$V_{h_{\mu\nu}}$ is well-defined.  
However, due to the zero, there appears the set of zero-norm states 
\be
{\cal X}_{\mu\nu}=\bigoplus_{k=0}^\infty 
\{\Lambda^h_{kp+\mu}, \Lambda^h_{kp+\mu+1}, \cdots,
\Lambda^h_{(k+1)p-1}\}.
\label{eq:null}
\ee
One can immediately show that, upon the relation 
$[kp+x]=(-)^k[x]$, $\parallel \chi \parallel^2=0$ for 
$\chi\in {\cal X}_{\mu,\nu}$. 
It is interesting to notice that the $V_{h_{\mu\nu}}$ actually 
form a discrete series parameterized by the two integers $\mu$ and
$\nu$. 
This is a characteristic feature of $\ncom$. 
Indeed this is not the case for the infinite dimensional 
representations of the classical algebra $\slc$ and of $\qslc$ 
with generic $q$ as well. 
It will be turned out that one of the parameter, say, $\nu$  
parameterizes the classical sector $V^{\slr}$  and the other, $\mu$,
is concerned with $V^{\qslc}_F$. 

Let $V_{\mu,\nu}:=V_{h_{\mu\nu}}$ and 
$\Lambda^{\mu\nu}_r:=\Lambda^{h_{\mu\nu}}_r$. 
The second problem we encounter in the construction of an irreducible 
highest weight module is that $X_\pm^p$ are nilpotent on the module 
$V_{\mu\nu}$, \ie, 
$(X_\pm)^p \Lambda =0$ for ${}^\forall \Lambda\in V_{\mu,\nu}$. 
Therefore one cannot move from a state to another state by acting 
$X_+$ or $X_-$ successively. 
Moreover, the relation $K^{2N}=1$ on $V_{\mu,\nu}$ 
indicates that one cannot measure the weight of a state completely. 
In order to remedy these situations, we should change the definition 
of $\qslc$ by adding new  generators, 
\be 
L_1:= -\frac{(-X_-)^p}{[p]!}, \quad L_{-1} := \frac{X_+^p}{[p]!}, 
\quad L_0 := \half \left[ \begin{array}{c} 2H+p-1 \\ p 
                          \end{array}\right]_q 
\ee
to the original ones, $X_+, X_-$ and $K=q^H$. 
The complete highest weight module $V_{\mu,\nu}$ is constructed by 
acting $X_+$ and $L_{-1}$ on the highest weight state 
$\Lambda^{\mu\nu}_0$ which is defined by 
the relations, $X_-\Lambda^{\mu\nu}_0=L_1\Lambda^{\mu\nu}_0=0$. 

The third problem, which we do not encounter in the classical case 
and the case when $q$ is not a root of unity, is that 
there is an infinite chain of submodules in $V_{\mu,\nu}$. 
That is to say, $V_{\mu,\nu}$ is no longer irreducible. 
Let us observe this characteristic feature briefly. 
First of all, one sees that the state $\Lambda^{\mu\nu}_\mu$ is 
a highest weight state because both $X_-$ and $L_1$ annihilate this state. 
Therefore a submodule exists on the state $\Lambda^{\mu\nu}_\mu$. 
Since this state has weight $h_{\mu\mu}+\mu=h_{-\mu\nu}$, 
the submodule  can be regarded as $V_{-\mu,\nu}$. 
Next one again finds a submodule $V_{\mu,\nu+2}$ on 
$\Lambda^{-\mu\nu}_{p-\mu}\in V_{-\mu,\nu}$. 
Repeating this procedure, one obtains the following chain of 
submodules in the original module $V_{\mu,\nu}$, 
\be 
V_{\mu,\nu}\supset V_{-\mu,\nu}\supset V_{\mu,\nu+2}\supset 
V_{-\mu,\nu+2}\supset\cdots \supset V_{\mu,\nu+2k}\supset 
V_{-\mu,\nu+2k}\supset\cdots . 
\ee
Then, irreducible highest weight module on the highest weight state 
$\Lambda^{\mu\nu}_0$, denoted by $V^{irr}_{\mu,\nu}$, 
is obtained by subtracting all the submodules, and it is obtained as 
\be 
V^{irr}_{\mu,\nu}=\bigoplus_{k=0}^\infty V^{(k)}_{\mu,\nu}, \quad 
V^{(k)}_{\mu,\nu}:= \{\Lambda^{\mu\nu}_{kp+s} \,\vert \,
s=0, 1, \cdots, \mu-1\}. \label{eq:irreducible}
\ee
It should be noticed that all the states subtracted in the procedure 
are the elements in the set ${\cal X}_{\mu,\nu}$ and, therefore, 
no zero norm state exists in $V^{irr}_{\mu,\nu}$, 
namely, 
\be
V_{\mu,\nu}^{irr}=V_{\mu,\nu}\backslash {\cal X}_{\mu,\nu}. 
\ee
We have now obtained  irreducible highest weight modules.  
Notice that every irreducible infinite dimensional highest weight 
representation is composed of an infinite number of blocks 
$V^{(k)}, k=0, 1, \cdots$, each of which contains finite number 
of states. 
This feature is the very origin of the fact stated in the
Theorem (\ref{eq:theorem}). To see this is the next task. 

Now we are the last stage to prove the Theorem. 
To complete the proof, we make some observations. 
In the following we will denote the level of a state by 
$kp+s$ where $s$ runs from $0$ to $\mu-1$ and $k=0, 1, \cdots$, 
instead of $r$, $r=0, 1, \cdots$.  

\vspace{.2cm}
\noindent
{\sc Observation} 1. \hspace{.2cm}On $V^{irr}_{\mu,\nu}$ 
\be 
\frac{X_\pm^{kp+s}}{[kp+s]!}=(-)^{\half k(k-1)p+ks}\frac{L_{\mp1}^k}{k!}
\frac{X_\pm^r}{[\,s\,]!}. 
\ee

\vspace{.2cm}
\noindent
{\sc Observation} 2. \hspace{.2cm}The sets of generators 
$\{X_+, X_-, K\}$ and $\{L_1, L_{-1}, L_0\}$ are mutually commutable 
on the irreducible highest weight module $V^{irr}_{\mu,\nu}$. 

\vspace{.2cm}
\noindent
These observations indicate that there exists a map 
$\rho\,:\, V^{irr}_{\mu,\nu}\rightarrow V^{inf}\otimes {\cal V}$, 
where ${\cal V}$ is the finite dimensional space composed of 
$\mu$ states. 
Further $\rho$ induces another map $\hat{\rho}\,:\, 
\qslc\rightarrow U^{inf}\otimes {\cal U}$, such that 
$\rho({\cal O}\Lambda)=\hat{\rho}({\cal O})\rho(\Lambda)$. 
In the second paper in Ref.\cite{MS}, 
such the isomorphisms $\rho$ and $\hat{\rho}$ have been obtained. 
In the following, we shall restrict the module $V^{irr}_{\mu,\nu}$ 
to the unitary irreducible representation. 
In that case, 
\be 
\rho(\Lambda^{\mu\nu}_{kp+s}) = (-)^{\half k(k-1)p+ks}
\lambda^\zeta_k \bigotimes \Psi^j_m, 
\label{eq:isomorphism}
\ee
and 
\begin{eqnarray}
\hat{\rho}(X_\pm) = {\bf 1}\otimes (\pm{\cal J}_\pm), &\quad&
\hat{\rho}(K) = {\bf 1} \otimes {\cal K}, \\
\hat{\rho}(L_{\pm1}) = (\mp G_{\pm1}) \otimes {\bf 1}, &\quad&  
\hat{\rho}(L_0) = G_0 \otimes {\bf 1},
\end{eqnarray}
where $\zeta=\nu/2, \, j=(\mu-1)/2$ and $m=-j+s$. 

\vspace{.2cm}
\noindent
{\sc Observation} 3. \hspace{.2cm} 
The actions of $\{{\cal J}_\pm, {\cal K}\}\in {\cal U}$ on the state 
$\Psi^j_m\in {\cal V}$ are calculated as  
\be 
{\cal J}_\pm \Psi^j_m = [j \pm m +1] \Psi^j_{m\pm1}, \quad 
{\cal K}\Psi^j_m = q^m \Psi^j_m, 
\ee 

\vspace{.2cm}
\noindent
{\sc Observation} 4. \hspace{.2cm}
The actions of $G_n, (n=\pm1, 0)\in U^{inf}$ 
on $\lambda^\zeta_k \in V^{inf}$ are as follows,  
\be
G_1 \lambda^\zeta_k = (2\zeta +k -1) \lambda^\zeta_{k-1}, 
\quad G_{-1} \lambda^\zeta_k = (k+1) \lambda^\zeta_{k+1}, 
\quad G_0 \lambda^\zeta_k = \left( \half\zeta+k\right)
\lambda^\zeta_k. 
\ee

\hspace{.2cm}
\noindent
The {\sc Observation} 3 shows that 
${\cal J}_\pm, {\cal K}$ satisfy the relations of 
$\qslc$, \ie, ${\cal U}=\qslc$  with the hermitian conjugations 
${\cal J}_\pm^\dagger = {\cal J}_\mp, {\cal K}^\dagger =
{\cal K}^{-1}$. 
Therefore, taking it into account that ${}^\forall\Psi^j_m$ 
has positive norm, ${\cal V}$ is a unitary finite dimensional 
representation of $\qslc$ with the highest weight $j$. 
We rewrite such the representation as ${\cal V}_j$, \ie, 
${\rm dim}.\,{\cal V}_j=2j+1$. 
On the other hand, {\sc Observation} 4 leads us to the following  
relations among $G_n, \, n=0, \pm1$, 
\be
[G_n. G_m]=(n-m) G_{n+m}, 
\ee 
and hermitian conjugations $G_{\pm1}^\dagger = G_{\mp1}, 
G_0^\dagger = G_0$. 
Therefore we can conclude that $U^{inf}$ is just the classical 
universal enveloping algebra of ${\rm su}(1,1)\cong\slr$, 
\ie, $U^{inf}=U\slr$ and $V^{inf}$ 
is the unitary highest weight representation of $\slr$ with highest 
weight $\zeta$. We denote the representation by $V^{cl}_\zeta$. 
We have now finished the proof of the theorem and 
obtained the important structure of $\ncom$ as 

\be 
V^{irr}_{\mu,\nu}\cong V^{cl}_\zeta \bigotimes {\cal V}_j.  
\label{eq:essential}
\ee 
In the Theorem, $V^{irr}_{\mu,\nu}$, $V^{cl}_\zeta$ and 
${\cal V}_j$ are denoted by $V^{\qslc}_I$, $V^{\slr}$ and 
$V^{\qslc}_F$, respectively.

%%%%%%%%%%%%%%%%                BIBLIOGRAPHY                 %%%%%%%%%%%%%%%%%%

\end{document}